\documentclass[12pt,a4paper]{article}
\pdfoutput=1

\usepackage{amsmath,amsfonts,stmaryrd,latexsym}
\usepackage{amssymb,a4wide,color,epsfig}

\usepackage{cite}
\usepackage{array}

\def \be  {\begin{equation}}
\def \ee  {\end{equation}}
\def \ba  {\begin{eqnarray}}
\def \ea  {\end{eqnarray}}
\def \bb  {\begin{thebibliography}}
\def \eb  {\end{thebibliography}}
\def \lab #1 {\label{#1}}

\newcommand\PT{\mathbb{PT}}

\newcommand\cM{\mathcal{M}}
\newcommand\cN{\mathcal{N}}

\newcommand\cZ{\mathcal{Z}}

\newcommand\MHV{\mathrm{MHV}}

\newcommand\tPhi{\widetilde{\Phi}}

\newcommand\C {\mathbb{C }}

\newcommand\rd{\mathrm{d}}
\newcommand\rD{\mathrm{D}}

\newcommand\la{\langle}
\newcommand\ra{\rangle}
\newcommand\del{\partial}

\newcommand{\bea}{\begin{eqnarray}\label}
\newcommand{\eea}{\end{eqnarray}}

\newcommand{\captionfonts}{\small}

\makeatletter  
\long\def\@makecaption#1#2{%
  \vskip\abovecaptionskip
  \sbox\@tempboxa{{\captionfonts #1: #2}}%
  \ifdim \wd\@tempboxa >\hsize
    {\captionfonts #1: #2\par}
  \else
    \hbox to\hsize{\hfil\box\@tempboxa\hfil}%
  \fi
  \vskip\belowcaptionskip}
\makeatother   

\begin{document}
  
\thispagestyle{empty}

\vskip1.5truecm
\begin{center}
\vskip 0.2truecm

 {\Large\bf
Gravity from Rational Curves
}
\vskip 1truecm

{\bf Freddy Cachazo and David Skinner\\}

\vskip 0.2truecm

{\it Perimeter Institute for Theoretical Physics,\\ 
	31~Caroline~St., Waterloo, Ontario N2L 2Y5, 
	Canada\\}
\vskip 1truecm

\end{center}

\centerline{\bf Abstract}
\medskip

This paper presents a new formula which is conjectured to yield all tree amplitudes in $\cN=8$ supergravity. The amplitudes are 
described in terms of higher degree rational maps to twistor space. The resulting expression has manifest $\cN=8$ supersymmetry and 
is manifestly permutation symmetric in all external states. It depends monomially on the infinity twistor that explicitly breaks conformal 
symmetry to Poincar{\' e}. The formula has been explicitly checked to yield the correct amplitudes for the 3-point 
$\overline\MHV$ and for the $n$-point MHV, where it reduces to an expression of Hodges. We have also carried out numerical checks of 
the formula at NMHV and N$^2$MHV level, for up to eight external states. 

\newpage


Twistors have proved very useful in the study of scattering amplitudes in $\cN=4$ SYM, since they make superconformal invariance 
manifest. General Relativity is of course not conformally invariant, and so a twistor description of gravity must involve a new structure that 
explicitly breaks the conformal invariance. The required structure is known as the {\it infinity twistor} $I$ and for flat 
space-time it takes the form
\be
	I^{ab} = \begin{pmatrix}
		\epsilon^{\dot\alpha\dot\beta} & 0\,\\
		0 & 0
		\end{pmatrix}
		\qquad\qquad\qquad
		I_{ab} := \epsilon_{abcd}I^{cd} = 
		\begin{pmatrix}
		\,0 & 0\\
		\,0 &\epsilon^{\alpha\beta}
		\end{pmatrix}\ .
\label{flatI}
\ee
The infinity twistor allows one to define a metric on space-time by
\be
	\frac{(1234)}{\la 12\ra\la34\ra}:=\frac{\epsilon(Z_1,Z_2,Z_3,Z_4)}{I_{ab}Z_1^aZ_2^b\ I_{cd}Z_3^cZ_4^d} =(x-y)^2\ ,
\label{metric}
\ee
where $x$ and $y$ are the points corresponding to the lines $12$ and $34$, respectively. This infinity twistor gets its name because in 
flat space it obeys $I_{ab}I^{ab}=0$ and so itself defines a point in conformally compactified space-time. This point plays a distinguished 
role and is $\imath^0$, the point at space-like infinity in the Penrose diagram. Twistor lines that intersect $I$ correspond to points that are 
null-separated from $\imath^0$, or in other words points on the null cone $\mathcal{I}^\pm$ at infinity. The infinity twistor should be 
thought of as providing a mass scale, so that equation~\eqref{metric} has dimensions of (length)$^2$.

The fact that gravitational scattering amplitudes break conformal invariance can be seen immediately once they are written in twistor 
space because, unlike the Yang-Mills amplitudes, they depend on the infinity twistor. For example, once transformed to twistor space, the 
3-point amplitudes become simply~\cite{ArkaniHamed:2009si,Mason:2009sa}
\be
\begin{aligned}
	\cM_{\overline\MHV}(\cZ_1,\cZ_2,\cZ_3) 
	&= I^{ab}\frac{\del}{\del Z^a_2} \bar\delta^{3|8}(\cZ_1;\cZ_2)\,\frac{\del}{\del Z_3^b}\bar\delta^{3|8}(\cZ_1;\cZ_3) \\
	\cM_\MHV(\cZ_1,\cZ_2,\cZ_3) &= I_{ab}Z_2^aZ_3^b\, \bar\delta^{2|8}(\cZ_1;\cZ_2,\cZ_3)
\end{aligned}
\label{gravity3}
\ee
where  $\bar\delta^{3|8}(\cZ_1,\cZ_2)$ has support only where $\cZ_1$ and $\cZ_2$ coincide in projective twistor space, and 
$\bar\delta^{2|8}(\cZ_1,\cZ_2,\cZ_3)$ has support only when the three projective twistors are collinear (see {\it e.g.}~\cite{Adamo:2011pv} 
for further details). Each of these gravitational amplitudes contains an irremovable infinity twistor and the absence of 
conformal invariance is just as manifest as its presence was in Yang-Mills.

\medskip

It is very instructive to examine the occurrence of the infinity twistor in arbitrary gravitational tree amplitudes. This can be done most
easily using the BCFW recursion relations, which in twistor space take the same form for gravity as for Yang-Mills. The standard 
spinor-helicity BCFW shift
\be
	\lambda_1 \to\lambda_1 + t\lambda_n\,,
	\qquad \tilde\lambda_n\to\tilde\lambda_n - t\tilde\lambda_1\,,
	\qquad \eta_n \to\eta_n - t\eta_1
\label{shBCFWshift}
\ee
becomes the shift
\be
	\cZ_1 \to \cZ_1 + t\cZ_n\,, \qquad \cZ_n \to \cZ_n
\label{twistorshift}
\ee
in twistor space, and the BCFW decomposition of an $n$-particle N$^{d-1}$MHV tree amplitude 
becomes~\cite{ArkaniHamed:2009si,Mason:2009sa}
\be
	\cM_{n,d}(1,\ldots,n) = \sum_{\begin{subarray}{c} n_L+n_R=n+2\\ d_L+d_R=d \end{subarray}} 
	\int\rD^{3|8}\cZ\,\frac{\rd t}{t} \,\cM_{n_L,d_L}(\cZ_1+t\cZ_n,\ldots,\cZ)\,\cM_{n_R,d_R}(\cZ,\cdots,\cZ_n)\ .
\label{gravBCFW}
\ee
The important point here is that both the BCFW shift and the recursion relation itself are fully superconformally invariant, as in 
Yang-Mills. The breaking of conformal invariance in arbitrary gravitational tree amplitudes is thus inherited entirely from its breaking in 
the 3-particle amplitudes that seed the recursion.

It is now straightforward to count how many powers of $I$ we expect any tree amplitude to contain. In fact, since
\be
	I^{ab}\frac{\del}{\del Z_1^a} \frac{\del}{\del Z_2^b} = \left[\frac{\del}{\del\mu_1}\,\frac{\del}{\del\mu_2}\right]
	\qquad\hbox{while}\qquad 
	I_{ab}Z_1^aZ_2^b = \la 12\ra\, ,
\ee
it actually makes sense to count powers of $[\ ,\,]$ and $\la\ ,\,\ra$ separately, since multiplication by $\lambda$ commutes with 
differentiation with respect to $\mu$. Consider first the power of $\la\ ,\,\ra$. The 3-point $\overline\MHV$ ($d=0$) and MHV ($d=1$) 
amplitudes contain 0 and 1 power of $\la\ ,\,\ra$, respectively. If we assume that for some $(d_L,d_R)$ the subamplitudes likewise 
contain $d_L$ and $d_R$ factors of $\la\ ,\,\ra$, then because BCFW recursion gives $d_L+d_R=d$, we learn by induction that an 
N$^{d-1}$MHV amplitude contains $d$ factors of $\la\,,\,\ra$ for all higher MHV levels, irrespective of the number of particles. Similarly,
the 3-point $\overline\MHV$ and MHV amplitudes contain 0 and 1 power of $[\ ,\,]$ respectively, if we assume that for some 
$(\{n_L,d_L\},\{n_R,d_R\})$ the subamplitudes contain $n_{L,R}-d_{L,R}-2$ powers of $[\ ,\,]$, then since
\be
	(n_L-d_L-2) + (n_R-d_R-2) = n-d-2
\ee
in BCFW, we induce that this is true for all $\{n,d\}$. To summarize, we have learned that when written in twistor space, the $n$-particle 
N$^{d-1}$MHV amplitude $\cM_{n,d}(\cZ_1,\ldots,\cZ_n)$ is a monomial of degree $d$ in $\la\ ,\,\ra$ and a monomial of degree 
$(n-d-2)$ in $[\ ,\,]$.

There are two useful consistency checks on these numbers. First, the total number of infinity twistors in an $n$-particle amplitude is 
$n-2$, irrespective of MHV level. Recalling that the infinity twistor behaves like a mass, this simply says that the kinematic structure of 
such amplitudes has dimensions of (mass)$^{n-2}$, and so needs to be accompanied by $n-2$ powers of the gravitational 
coupling $\kappa\sim \sqrt{G_{\rm N}}$ so that the amplitude can be a complex number. This is exactly as expected from expanding the 
Einstein-Hilbert action $S_{\rm EH}= \frac{1}{\kappa^2}\int *_g \,R(g)$ around some background and normalizing the fluctuations to 
have a canonical kinetic term. Secondly, under parity conjugation an $n$-particle N$^k$MHV amplitude becomes an $n$-particle N$^{n-k-4}$MHV amplitude, or in other words
\be
	d \leftrightarrow (n-d-2)\,.
\label{parity}
\ee
Thus the number of irremovable angle and square brackets is interchanged under parity, with their total number remaining the same. 
This reflects the fact that we could equally construct the parity conjugate amplitude by using the dual twistor space, where $\la\ ,\,\ra$ 
becomes the differential operator $I_{ab}\frac{\del}{\del W_{1a}}\frac{\del}{\del W_{2b}}$ while $[\ ,\,]$ becomes multiplication by 
$I^{ab}W_{1a}W_{2b}$.

Knowledge of the amplitudes' dependence on the infinity twistor provides important insight into their general twistor space structure. 
In Hodges' beautiful reformulation of MHV amplitudes~\cite{Hodges:2011wm,Hodges:2012ym}, one obtains $n-3$ powers of $I^{ab}$ by 
starting with an $n\times n$ matrix, each of whose entries are linear in the infinity twistor. The matrix has rank $n-3$, so its reduced 
determinant provides the correct dependence on $I^{ab}$. This suggests a natural extension to non-MHV amplitudes, which we now 
present.


\bigskip

The main claim of this paper is that the complete $n$-particle classical S-matrix of  $\cN=8$ supergravity is given by the formula
\be
	\cM_n = \sum_{d=0}^\infty\int \frac{\prod_{{\rm a}=0}^d\rd^{4|8}\cZ_{\rm a}}{{\rm vol\,GL(2;\C)}}\ {\det}'(\Phi)\,{\det}'(\tPhi)\, 
	\prod_{i=1}^n \rd^2\sigma_i\, \delta^2(\lambda_i-\lambda(\sigma_i))\exp\, \llbracket \mu(\sigma_i)\tilde\lambda_i\rrbracket
\label{N=8}
\ee
whose ingredients we now explain. Firstly, as in the twistor-string~\cite{Witten:2003nn,Berkovits:2004hg}, $\cZ:\Sigma\to\PT$ is a degree 
$d$ holomorphic map from a Riemann sphere to $\cN=8$ supertwistor space, described by degree $d$ polynomials
\be
	\cZ^I(\sigma) = \sum_{{\rm a}=0}^d \cZ^I_{\rm a}(\sigma^{\underline 1})^{\rm a}(\sigma^{\underline 2})^{d-{\rm a}}
\ee
in the homogeneous coordinates $\sigma^{\underline\alpha} = (\sigma^{\underline 1},\sigma^{\underline 2})$ on $\Sigma$. We integrate 
over the space of such maps upto an overall scale. Note that unlike the case of $\cN=4$ supersymmetry, neither $\cN=8$ twistor space 
itself nor the projective space of holomorphic maps is  Calabi-Yau, and the integration measure here scales like the 
$-4(d+1)^{\rm th}$ power of $\cZ$. This scaling is compensated by the rest of the integrand.

The factors of
$$
	\delta^2(\lambda_i - \lambda(\sigma_i))\exp\,\llbracket\mu(\sigma_i)\tilde\lambda_i\rrbracket
$$
are the wavefunctions of the external states, which we take to be twistor representatives of momentum eigenstates. We define
$\llbracket\mu\tilde\lambda\rrbracket := \mu^{\dot\alpha}\tilde\lambda_{\dot\alpha} + \chi^A\eta_A$ to include the fermionic components.
The insertions points of these vertex operators on $\Sigma$ are integrated over up to the usual SL$(2;\C)$ redundancy.

The important content of~\eqref{N=8} is the operators ${\det}'(\Phi)$ and ${\det}'(\tPhi)$. Let $\tPhi$ be the symmetric $n\times n$ matrix whose entries 
are the operators
\be
\begin{aligned}
	\tPhi_{ij} &= \frac{[ij]}{(ij)}  
	\hspace{3.5cm}  \hbox{for } i\neq j,\\
	\tPhi_{ii} &= -\sum_{j\neq i}  \tPhi_{ij} \prod_{{\rm a}=0}^d\frac{(j\,p_{\rm a})}{(i\,p_{\rm a})} \ ,
\end{aligned}
\label{tphidef}
\ee
where $(ij):=\epsilon_{\underline{\alpha}\underline{\beta}}\sigma_i^{\underline\alpha}\sigma_j^{\underline\beta}$ is the 
SL$(2;\C)$-invariant inner product of worldsheet coordinates.  Notice that each element of $\tPhi$ contains a single factor of the infinity 
twistor $I^{ab}$.

In writing the diagonal term, we have picked $d+1$ arbitrary points $p_0,\ldots,p_{d}\in\Sigma$. We first show that $\tPhi$ is independent 
of this choice. Since $\tPhi_{ii}$ is clearly symmetric under arbitrary permutations of the $p_{\rm a}$, it suffices to show independence 
under a change in any one of the $p_{\rm a}$, say $p_0\to q_0$. We have
\be
\begin{aligned}
	\Delta\tPhi_{ii} &= -\sum_{j\neq i}  \frac{[ij]}{(ij)} \left(\frac{(jq_0)}{(iq_0)}-\frac{(jp_0)}{(ip_0)}\right)
	\prod_{{\rm a}=1}^d\frac{(jp_{\rm a})}{(ip_{\rm a})} \\
	&=-\sum_{j=1}^n \frac{[ij]\,(p_0q_0)}{(iq_0)} \prod_{{\rm a}=1}^d\frac{(jp_{\rm a})}{(ip_{\rm a})} 
\end{aligned}
\ee
where in going to the second line we cancelled the factor $(ij)$ and then extended the sum over $j$ to all $n$ particles. This expression 
now vanishes on the support of the $\delta$-functions obtained by integrating out the $\mu$-components of the map $\cZ$ using the 
exponentials in the wavefunctions\footnote{This fact, which was also used in~\cite{Cachazo:2012uq}, was also a big clue leading to this formula.}.

It follows immediately that $\tPhi$ has rank $(n-d-2)$ for
\be
\begin{aligned}
	\sum_{j=1}^n\tPhi_{ij}\sigma_j^{\underline{\alpha}_1}\cdots\sigma_j^{\underline{\alpha}_{d+1}} 
	&= \sum_{j\neq i}  \frac{[ij]}{(ij)} \left\{\sigma_j^{\underline{\alpha}_1}\cdots\sigma_j^{\underline{\alpha}_{d+1}} 
	- \sigma_i^{\underline{\alpha}_1}\cdots\sigma_i^{\underline{\alpha}_{d+1}}\prod_{a=0}^d\frac{(jp_a)}{(ip_a)}\right\} \\
	&=\ 0\ ,
\end{aligned}
\label{nullvectors}
\ee
because we have just shown that $\tPhi$ is independent of the choice of $p_{\rm a}$, so we can always choose them to coincide with the 
choice of components on each of the $(d+1)$ $\sigma$s, whereupon the two terms in braces immediately cancel. Thus, as an operator 
on the external wavefunctions, $\sum_j\tPhi_{ij}P_{d+1}(\sigma_j)$ vanishes for an arbitrary polynomial $P_{d+1}(\sigma)$ of degree 
$d+1$.

Because $\tPhi$ has rank $n-d-2$, its determinant vanishes to $(d+2)^{\rm th}$ order. As in Hodges' MHV formula, let 
$|\tPhi_{\rm red}|$ denote the determinant of the reduced matrix $\tPhi_{\rm red}$ obtained by removing any $(d+2)$ rows and, 
independently, any $(d+2)$ columns from $\tPhi$.  $|\tPhi_{\rm red}|$ itself is clearly symmetric under arbitrary 
permutations of states whose rows and columns both remain, since pick up a minus sign from exchanging their rows and another 
from exchanging their columns, it clearly depends on the choice of deleted rows and columns. However, a straightforward generalization 
of an argument used by Hodges (\!\!\cite{Hodges:2012ym}, at equation 8) shows that the ratio
\be
	{\det}'(\tPhi) :=\frac{|\tPhi_{\rm red}|}{|\tilde r_1\cdots \tilde r_{d+2}|\,|\tilde c_1\cdots \tilde c_{d+2}|}
\label{dettildephidef}
\ee
is symmetric under arbitrary permutations of all $n$ points. In this ratio, $|\tilde r_1\cdots \tilde r_{d+2}|$ denotes the Vandermonde 
determinant 
\be
	|\tilde r_1\cdots \tilde r_{d+2}| = \prod_{\begin{subarray}{c} i,j \in \{{\rm removed}\}\\ i<j \end{subarray}}(ij)
\label{vdmremove}
\ee
made from all possible combinations of the worldsheet coordinates corresponding to the deleted rows; 
$|\tilde c_1\cdots \tilde c_{d+2}|$ is the same Vandermonde determinant, but for the deleted columns.

\medskip

The remaining ingredient ${\det}'(\Phi)$ is constructed similarly. Let $\Phi$ denote the symmetric $n\times n$ matrix whose entries are
\be
\begin{aligned}
	\Phi_{ij} &= \frac{\la ij\ra}{(ij)}
	\hspace{5.5cm}  \hbox{for } i\neq j,\\
	\Phi_{ii} &= -\sum_{j\neq i} \left\{\Phi_{ij}\,
	\frac{\prod_{k\neq i}(ik)}{\prod_{l\neq j}(jl)}
	\prod_{{\rm a}=0}^{n-d-2}\frac{(jp_{\rm a})}{(ip_{\rm a})} \right\}\ ,
\end{aligned}
\label{phidef}
\ee
Each entry of $\Phi$ is linear in the infinity twistor $\la\ ,\,\ra$. To define this matrix we have picked a further 
$n-d-1$ points $p_{\rm a}\in\Sigma$ and we again wish to show that our matrix is independent of these choices. As before, upon changing any one of these points a Schouten identity gives
\be
	\Delta\Phi_{ii} \propto -\sum_{j=1}^n \la ij\ra \frac{(ij)\prod_{a=0}^{n-d-3}(j p_{\rm a})}{\prod_{k\neq j}(jk)}
\label{DeltaPhi}
\ee
where we have removed an overall factor\footnote{Note that the bracket $(ij)$ cancels in the ratio of products in $\Phi_{ii}$. This is the 
origin of this factor in~\eqref{DeltaPhi}.}. It was shown~\cite{Roiban:2004yf,Witten:2004cp} that the right hand side of~\eqref{DeltaPhi}  
vanishes on the support of the $\delta$-functions inside~\eqref{N=8}, so $\Phi$ is indeed independent of the choice of points. This again 
implies that $\Phi$ has rank $d$, with kernel spanned by the $n$-dimensional vectors whose $j^{\rm th}$ entry is $\sigma^{\underline\alpha_1}_j\cdots\sigma^{\underline\alpha_{n-d-1}}_j |1\cdots \widehat\jmath\cdots n|$, 
where $|1\cdots \widehat\jmath\cdots n|$ denotes the Vandermonde determinant of all points except $j$. The different choices for the
$\underline\alpha$ indices span the $(n-d)$-dimensional kernel.

Now let $|\Phi_{\rm red}|$ denote the determinant of the $d\times d$ reduced matrix obtained by deleting any $n-d$ rows and any $n-d$ 
columns from $\Phi$. Then
\be
	{\det}'(\Phi):=\frac{|\Phi_{\rm red}|}{|r_1\cdots r_d|\,|c_1\cdots c_d|}
\label{detphidef}
\ee
is independent of the choice of rows and columns. Here, in contrast to $\tPhi$,  $|r_1\cdots r_d|$ is the Vandermonde determinant
\be
	|r_1\cdots r_d| = \prod_{\begin{subarray}{c} i,j \in \{{\rm remain}\}\\ i<j \end{subarray}}(ij)
\label{vdmremain}
\ee
over the rows that {\it remain} in the reduced matrix $\Phi_{\rm red}$, and $|c_1\cdots c_d|$ is similar for the remaining columns. (When 
$d\leq1$, no such Vandermonde determinants can be formed. A careful treatment of Hodges' argument for the independence shows that 
in these two cases, the Vandermonde factors should be unity.)

All the dependence on the infinity twistor lies in ${\det}'(\Phi)$ and ${\det}'(\tPhi)$ -- if we scatter twistor eigenstates 
then the factors of $I$ from these determinants will be the only appearance of the infinity twistor in the amplitude. These operators thus 
make manifest the dependence on the infinity twistors deduced from BCFW recursion. 

\medskip

We have checked analytically that~\eqref{N=8} correctly reproduces the 3-particle $\overline\MHV$ amplitude, all $n$-particle MHV 
amplitudes. We also checked analytically that the 5 particle NMHV and 6 particle N$^2$MHV amplitudes agree with the formula 
of~\cite{Cachazo:2012da}. Finally, we have checked numerically that~\eqref{N=8} gives the correct six and seven particle NMHV 
amplitudes and the correct eight particle N$^2$MHV amplitude.

\medskip

To conclude, we have presented a new formulation of all tree amplitudes in $\cN=8$ supergravity as an integral over spaces of 
holomorphic maps into twistor space. The formula has manifest $\cN=8$ Poincar{\'e} supersymmetry is permutation 
invariant in all external states. It is based on degree $d$ holomorphic maps to $\cN=8$ twistor space and seems to augur the existence of 
a twistor-string for gravity, perhaps along the lines suggested in~\cite{Adamo:2012nn}. A somewhat similar formulation -- derived 
independently -- was obtained recently by Geyer and one of us~\cite{Cachazo:2012da}. We believe the new formula is a great 
improvement, both in terms of calculational simplicity and clarity of structure. We hope it reflects the underlying worldsheet theory more 
directly.

\vspace{1cm}

\noindent{\bf\Large Acknowledgments}

\bigskip

\noindent  It is a pleasure to thank Yvonne Geyer and Lionel Mason for very helpful discussions. This work is supported by the Perimeter 
Institute for Theoretical Physics. Research at the Perimeter Institute is supported by the Government of Canada through Industry Canada 
and by the Province of Ontario through the Ministry of Research $\&$ Innovation. The work of FC is supported in part by the NSERC of 
Canada and MEDT of Ontario.


\begin{thebibliography}{99}
\providecommand{\href}[2]{#2}

\bibitem{ArkaniHamed:2009si}
N.~Arkani-Hamed, F.~Cachazo, C.~Cheung and J.~Kaplan, {The S-Matrix in Twistor Space},
 {\em JHEP} {\bf 03} (2010) 110, 
  [\href{http://arxiv.org/abs/0903.2110}{{\tt arXiv:0903.2110}}].

\bibitem{Mason:2009sa}
L.~Mason and D.~Skinner, {Scattering Amplitudes and BCFW Recursion in Twistor Space}, 
  {\em JHEP} {\bf 01} (2010) 064,
  [\href{http://arxiv.org/abs/0903.2083}{{\tt arXiv:0903.2083}}].

\bibitem{Adamo:2011pv}
T.~Adamo, M.~Bullimore, L.~Mason, and D.~Skinner, {Scattering Amplitudes and Wilson Loops in Twistor Space},
 {\em J.~Phys.} {\bf A44} (2011) 454008,
  [\href{http://arxiv.org/abs/1104.2890 [hep-th]}{{\tt arXiv:1104.2890}}].

\bibitem{Hodges:2011wm}
A.~Hodges, {New Expressions for Gravitational Scattering Amplitudes},
  [\href{http://arxiv.org/abs/1108.2227}{{\tt arXiv:1108.2227}}].
  
\bibitem{Hodges:2012ym}
A.~Hodges, {A Simple Formula for Gravitational MHV Amplitudes},
  [\href{http://arxiv.org/abs/1204.1930}{{\tt arXiv:1204.1930}}].

\bibitem{Witten:2003nn}
E.~Witten, {Perturbative Gauge Theory as a String Theory in Twistor Space},
  {\em Commun. Math. Phys.} {\bf 252} (2004) 189-258, [\href{http://arxiv.org/abs/hep-th/0312171}{{\tt hep-th/0312171}}].

\bibitem{Berkovits:2004hg}
N.~Berkovits, {An Alternative String Theory in Twistor Space for $\mathcal{N}=4$ Super-Yang-Mills},
 {\em Phys. Rev. Lett.} {\bf 93} (2004) 011601, [\href{http://arxiv.org/abs/hep-th/0402045}{{\tt hep-th/0402045}}].

\bibitem{Cachazo:2012uq}
F.~Cachazo, {Fundamental BCJ Relation from the Connected Formulation},
 [\href{http://arxiv.org/abs/1206.5970}{{\tt arXiv:1206.5970}}].

\bibitem{Roiban:2004yf}
R.~Roiban, M.~Spradlin, and A.~Volovich, {On the Tree-Level S-Matrix of Yang-Mills Theory}, 
 {\em Phys. Rev.} {\bf D70} (2004) 026009,
  [\href{http://arxiv.org/abs/hep-th/0403190}{{\tt hep-th/0403190}}].

\bibitem{Witten:2004cp}
E.~Witten, {Parity Invariance for Strings in Twistor Space}, 
{\em Adv. Theor. Math. Phys.} {\bf 8} (2004) 779--796,
  [\href{http://arxiv.org/abs/hep-th/0403199}{{\tt hep-th/0403199}}].

\bibitem{Adamo:2012nn}
 T.~Adamo and L.~Mason, {Einstein Supergravity Amplitudes from Twistor-String Theory},
 {\em Class. Quant. Grav.} {\bf 29} (2012) 145010,
  [\href{http://arxiv.org/abs/1203.1026}{{\tt arXiv:1203.1026}}].

\bibitem{Cachazo:2012da}
F.~Cachazo and Y.~Geyer, {A `Twistor String' Inspired Formula for Tree-Level Scattering Amplitudes in $\mathcal{N}=8$ Supergravity},
  [\href{http://arxiv.org/abs/1206.6511}{{\tt arXiv:1206.6511}}].

\end{thebibliography}
\end{document}